# Snowmass Summer Study 2013
# White Paper for Cosmic Frontier (CF) Sub-Group

## *Dedicated Indirect Searches for Dark Matter Using Antideuterons*

C. Hailey[1], T. Aramaki[1], P. von Doetinchem[2], and R. A. Ong[3]

[1]Dept. of Physics, Columbia University, New York, NY 10027, USA

[2]Space Science Laboratory, University of California, Berkeley, CA 94720, USA

[3]Dept of Physics and Astronomy, University of California, Los Angeles, CA 90095, USA

### 1.) Basic Idea of Antideuteron Searches

About a decade ago it was pointed out that antideuterons produced in WIMP-WIMP annihilations (the "primary" antideuterons) offered a potentially attractive signature for cold dark matter (CDM) (1, hereafter DFS). The reason is that the flux of primary antideuterons is fairly flat in the ~ 0.1 – 1.0 GeV/n energy band, while the "secondary/tertiary" antideuterons, those produced in cosmic ray interactions in the interstellar medium (secondaries) and subsequent reprocessing (tertiaries), have fluxes which sharply decrease with decreasing energy. The lower antideuteron background results because of the higher cosmic ray energy required to create an antideuteron, compared to an antiproton, combined with a cosmic ray spectrum steeply falling with energy. In addition, the collision kinematics disfavors the formation of low-energy antideuterons.

Despite the low astrophysical background and a signature which is rather generic in many "beyond-the-Standard-Model" models, there has been no dedicated, optimized search for antideuterons. An upper limit on antideuterons was obtained by the BESS experiment (2), but it is three orders of magnitude higher than the interesting range for dark matter searches. The AMS experiment on ISS has sensitivity for antideuterons (3), but, overall, this is a rather modest assault given the ~20 operating or planned experiments to directly detect WIMPS through recoil interactions with target nuclei (4). Just such an optimized search for antideuterons has been proposed – the General Antiparticle Spectrometer experiment (GAPS) (5), and it recently had a successful prototype flight (6). The sensitivities of GAPS and AMS have generally driven the theoretical discussions about long and near term capabilities. Thus the discussion below is focused on several variants of these experiments in order to illustrate the reach of antideuteron searches.

## 2.) Model Predictions and Antideuteron Searches

Figure 1 shows antideuteron fluxes along with secondary/tertiary backgrounds expected for three generic, benchmark DM candidates. These are the neutralino, the lightest supersymmetric partner (LSP) from supersymmetric models, a Kaluza-Klein particle (LKP) and a 5D-warped GUT Dirac neutrino (LZP) (7). Sensitivities for GAPS long duration balloon flights (LDB for 105 days and LDB+ for 210 days) and AMS for 5 years of operation are also shown in Figure 1 (8). The experiment sensitivities illustrate three important points: (i) Antideuteron searches, as DFS pointed out, can have detection thresholds well above astrophysical backgrounds, especially in the low-energy band, (ii) there are models in which antideuteron searches can potentially detect CDM, as opposed to obtaining upper limits, and (iii) operational or planned antideuteron searches can improve on the "first generation" BESS results by two to three orders of magnitude.

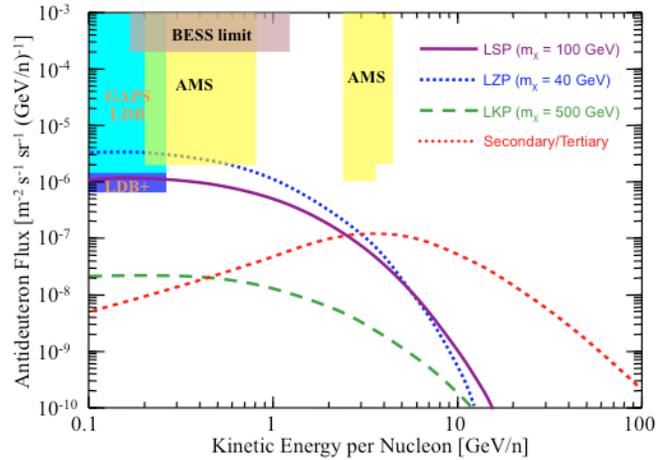

**Figure 1:** LSP, LKP and LZP models (7) along with the antideuteron background (8) and the sensitivity of GAPS (8), BESS (2) and AMS for 5 years of operation (8).

Figure 2 shows a recent set of models overlaid on the sensitivity for a dedicated antideuteron search experiment under various scenarios. This experiment can be executed from a balloon and have observations of duration from a few to ten months. This plot illustrates some of the opportunities as well as challenges of antideuteron searches. An ensemble of supersymmetric (SUSY) model parameters is shown, all yielding the same neutralino-annihilation cross-section and mass. To the right of the vertical green line are results from a low-energy minimal SUSY model, and to the left are shown results from a low mass non-universal gaugino model. The latter admits light neutralinos, which have recently been argued to provide a candidate for the controversial DAMA/LIBRA, CoGENT and CDMS II signals (10,11,12). The red dots indicate parameter space in the WMAP preferred density range, while the blue dots correspond to models in which the thermally-generated neutralinos are subdominant. The gray models are ruled out by antiproton searches. It has been argued that it is important to search the entire parameter space, not just the WMAP preferred range, since there are many mechanisms to under-produce WIMPS in the early universe and still have them detectable today (13). The solid and dashed lines indicate the sensitivity of the antideuteron search for the case of nominal (solid) and maximal (dashed) propagation models. There are several points illustrated by this figure. Firstly, antideuteron searches are quite sensitive to models with low energy neutralinos, and they maintain sensitivity up high neutralino masses. Secondly, the plot illustrates both the opportunity and the curse of an antideuteron search – antideuteron propagation and production uncertainties. This is discussed more below, but we note here that the reach into parameter space for primary antideuterons is sensitive to details of the propagation (and less so the production) of the antideuterons in the interstellar medium, which are still poorly constrained. Thus, for the best case uncertainties, antideuterons provide a deep reach into parameter space. Conversely, compounding the most pessimistic values of the uncertainties can lead to a reach, in a short balloon observation, that is not much better than has been obtained for antiproton searches.

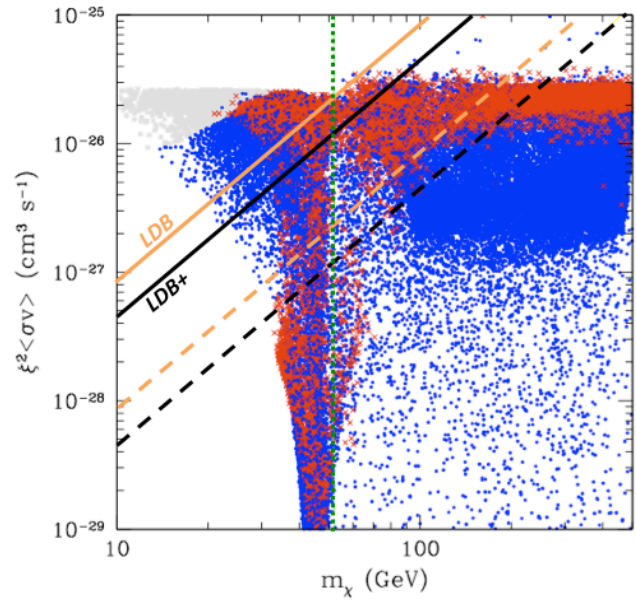

**Figure 2:** Sensitivity of a dedicated antideuteron experiment (diagonal lines) for long-duration balloon (LDB/LDB+) flights; green vertical line separates low mass non-universal gaugino SUSY models (left) from MSSM models (9). Solid (dashed) lines for the sensitivity correspond to median (high) antideuteron propagation models. The colors for the points are discussed in the text.

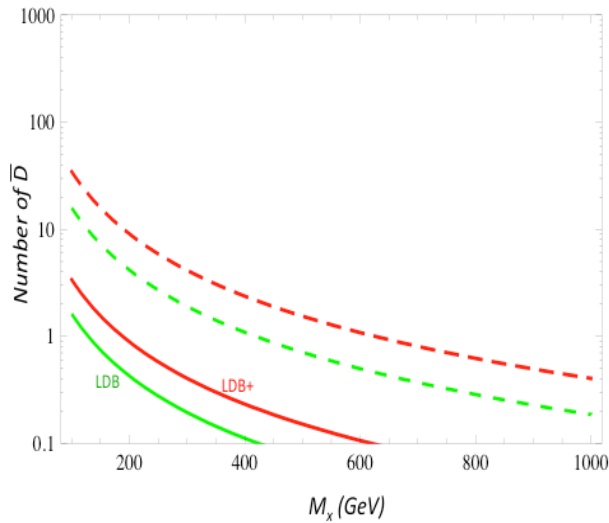 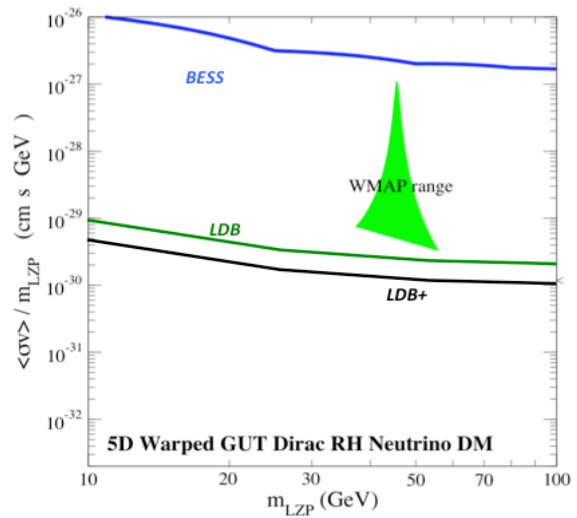

**Figure 3:** Detected antideuterons vs. neutralino mass in the gluon-gluon channel for GAPS LDB/LDB+ (14). Solid (dashed) lines correspond to median (high) antideuteron propagation models.

**Figure 4:** LZP sensitivities of GAPS with model from (7). Green is WMAP allowed density range.

Figure 3 shows a typical range of antideuteron counts as a function of neutralino mass for the gluon-gluon decay channel that might be realized in a balloon experiment (14). Figure 4 shows that antideuteron searches for the LZP can "clean out" the WMAP-preferred density range (7). Figures 5 and 6 illustrate antideuteron searches as discussed in several recent papers. Figure 5 shows the reach of these searches for a massive DM particle with mass in the ~5-20 TeV range (lower masses run afoul of antiproton limits) and large annihilation cross-section (15). This candidate was proposed to be consistent with the PAMELA observations of an excess in the positron fraction, but none in the antiproton fraction. Similar to the neutralino case of figure 2, the most favorable propagation models yield detections up to very high masses, while the unfavorable propagation cases permit detection of only the lightest DM candidates. And finally the case of the decaying gravitino is shown in figure 6 (16), with some typical propagation uncertainties superimposed. The plot illustrates that the decay of lighter gravitinos can be detected through the antideuteron channel if their mass is not too large. The decay

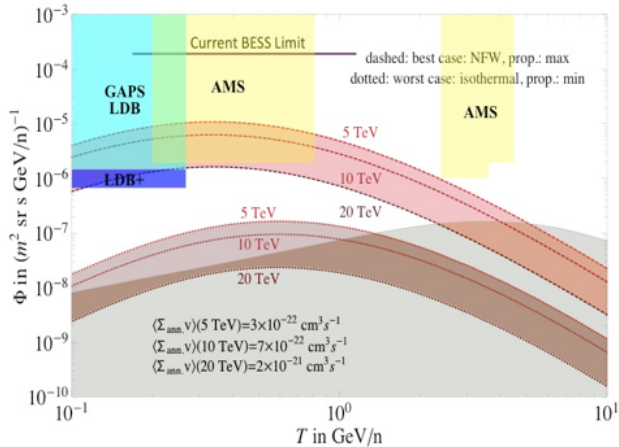 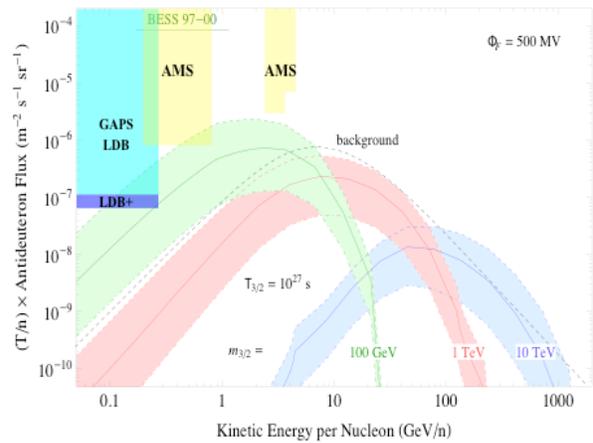

**Figure 5:** Sensitivity of antideuteron searches for massive neutralino annihilation. Shaded bands show different neutralino masses (5 TeV-20 TeV) for each propagation model (15).

**Figure 6:** Sensitivity of antideuteron searches for decaying gravitinos at an interesting decay rate. Shaded bands show the propagation uncertainty for each gravitino mass (16).

rate for which these plots were produced is not ruled out by any other experiment.

### 3.) Synergy and Complementarity with Other Search Methods

Many papers argue that a suite of experiments is essential to discover dark matter and constrain its properties. It was recognized early that since direct detection (and neutrino) experiments probe the WIMP scattering cross-section and indirect detection experiments probe the annihilation cross-section, joint observations work together to better constrain model parameters. Overlap of experiments and observations within model parameter spaces can increase confidence in the interpretation of signals, especially in the case of poorly understood astrophysical backgrounds. Antideuteron searches are timely in this regard. Because antideuteron searches are very sensitive to SUSY models which provide low mass neutralino candidates, their overlap with direct detection experiments is potentially of great utility. Similarly, antideuteron searches will provide overlap with Fermi. Fermi observations of dwarf spherical galaxies are beginning to place lower limits on neutralino masses (~ few tens of GeV) in the b-bbar and $\tau^+\tau^-$ decay channel (17), while antideuteron observations can provide constraints in the former channel, as well as $W^+W^-$. Antideuteron searches are generally sensitive to much higher neutralino masses than direct detection searches with massive target nuclei. Therefore antideuteron searches can complement the direct searches by providing reach at these higher masses.

### 4.) Antideuterons – the Fine Print

As mentioned above, there are substantial uncertainties in predictions of the primary antideuteron flux. The dominant uncertainty in the primary flux is due to propagation uncertainties. During the next years the AMS high precision observations will tightly constrain the cosmic ray propagation parameters. However, some degeneracy in the parameters will most likely remain. There is also an uncertainty due to the halo model employed, but the antideuteron production is averaged over the halo and over fairly long scale lengths, so whether or not a cored or uncored halo profile is used is somewhat reduced in importance. Recently, increasing attention has been paid to details of the hadronization process. In the coalescence model, the antineutron and antiproton combine when their momentum difference is less than a critical value – the coalescence momentum, $p_o$. But since $p_o$ is smaller than the QCD phase transition temperature, there is a strong sensitivity to the hadronization model. Recent studies emphasize the need for event-by-event determination of the production rates and are probing the sensitivities when different hadronization models are employed (18,19,20). In addition, $p_o$ is experimentally determined; its value is not well constrained and antideuteron production goes like $p_o^3$. So while theoretical progress is being made, the best way to pin down the production uncertainty is unquestionably to make a good accelerator measurement of $p_o$, given the considerable discrepancy in the various measurements. One possible source of uncertainty, which only drives up the expected primary flux, is due to the boost factor. It was previously fashionable to consider boost factors up to 100 times or more, but recent simulations seem to have definitively settled this issue, with boost factors in the range of 1-10 being the largest allowed (21). However the sensitivity curves presented here assume no boosting effect (i.e., boost factor 1), and a factor of 2 or 3 in boost would provide very substantial reach into discovery space for a number of models presented here.

Similar observations pertain to the secondary/tertiary background, except that in this case it is the production uncertainties that dominate (9). Most of the secondary/tertiary antideuterons are produced in the Galactic disk, so the propagation is more local and thus less important.

At any rate, for the antideuteron searches the nominal or optimistic propagation and production parameters are very promising for dark matter searches. The most pessimistic numbers would make prospects for detection problematic.

### 5.) The Path Forward: How Many Experiments?

An analogy to underground direct searches is appropriate. The direct searches require long integration times (months to years), are looking for just a handful of events, depending on model, and are dominated by internal backgrounds. This is much the same case for antideuteron searches, especially low-energy antideuteron searches. The direct search community has addressed this challenge by building multiple experiments. Even when these experiments have comparable sensitivities, they exploit different targets, their instrumental backgrounds are not identical and they often employ completely different approaches to suppressing this background.

Much the same considerations apply to the leptonic and photon-based indirect searches for dark matter. Only by applying a variety of instruments with different origins and sensitivities to backgrounds (eg. Leptons: PAMELA, Fermi, ATIC etc.; photons: Fermi, HESS, VERITAS; WMAP etc.) is it possible to disentangle a true dark matter signal from a confounding internal or external background source. In the near future the antideuteron search will exclusively rely on AMS, a multi-purpose cosmic ray detector on the ISS, and the General AntiParticle Spectrometer (GAPS), which is a dedicated low-energy antideuteron detector in the design phase planned to fly aboard high altitude long duration balloons from Antarctica.

AMS and GAPS have mostly complementary kinetic energy ranges, but also some overlap in the interesting low energy region, which allows the study of both a large energy range, confirming the potential signals, and the best chance for controlling the systematic effects. Another very important virtue comes from the different detection techniques of both experiments. AMS follows the principle of typical particle physics detectors. Particles are identified by analyzing the event signatures of different subsequent subdetectors, also using a strong magnetic field. The GAPS detector will consist of several planes of Si(Li) solid state detectors and a surrounding time-of-flight system. The antideuterons will be slowed down in the Si(Li) material, replace a shell electron and form an excited exotic atom. The atom will be deexcited by characteristic X-ray transitions and will end its life by annihilation with the nucleus producing a characteristic number of protons and pions. The approach of two independent experiments has certainly been successfully employed by the CDMS and XENON collaborations, and is likely to be vital in providing confidence in any potential primary antideuteron detection. The GAPS antideuteron search is thus highly desirable. Unless such an experiment is begun soon, confirmation of an AMS detection would be a long while forthcoming. Similarly a promising method to search for dark matter will stagnate if a non-detection by AMS is not quickly followed by a more sensitive experiment.

+ The sensitivity plots in the figures of this white paper have been modified from the published versions to update to the most recent calculations of GAPS sensitivity, and calculated backgrounds.

++ The sensitivity lines in the figures refer to the antideuteron flux for which a detection indicate a ~99% confidence that the event was not a secondary/tertiary antideuteron.